\documentclass[sigplan,screen]{acmart}

\usepackage[]{hyperref}
\usepackage{bm}
\usepackage{amsmath}
\usepackage{algorithm}  
\usepackage{algorithmicx}  
\usepackage{algpseudocode}
\usepackage{mathtools}
\usepackage{xspace}
\usepackage{multirow}
\usepackage{bm}
\newcommand{\nickname}{\textit{FlashOverlap}\xspace}
\newcommand{\hk}[1]{\textcolor{black}{#1}}

\AtBeginDocument{%
  }

\setcopyright{acmlicensed}

\acmYear{2026}\copyrightyear{2026}
\setcopyright{cc}
\setcctype[4.0]{by}
\acmConference[EUROSYS '26]{European Conference on Computer Systems}{April 27--30, 2026}{Edinburgh, Scotland Uk}
\acmBooktitle{European Conference on Computer Systems (EUROSYS '26), April 27--30, 2026, Edinburgh, Scotland Uk}
\acmDOI{10.1145/3767295.3769370}
\acmISBN{979-8-4007-2212-7/26/04}

\begin{document}

\title{Efficient and Adaptable Overlapping for Computation and Communication via Signaling and Reordering}

\author{Ke Hong}
\affiliation{
  \institution{Tsinghua University, Infinigence-AI}
  \country{}
}

\author{Xiuhong Li}
\affiliation{%
  \institution{Infinigence-AI}
  \country{}
}

\author{Minxu Liu}
\affiliation{%
  \institution{Infinigence-AI}
  \country{}
}

\author{Qiuli Mao}
\affiliation{%
  \institution{Infinigence-AI}
  \country{}
}

\author{Tianqi Wu}
\affiliation{%
 \institution{Tsinghua University, Infinigence-AI}
 \country{}}

\author{Zixiao Huang}
\affiliation{%
  \institution{Tsinghua University, Infinigence-AI}
  \country{}}

\author{Lufang Chen}
\affiliation{%
  \institution{Infinigence-AI}
  \country{}}

\author{Zhong Wang}
\affiliation{%
  \institution{Tsinghua University}
  \country{}}

\author{Yichong Zhang}
\affiliation{%
  \institution{Tsinghua University}
  \country{}}

\author{Zhenhua Zhu}
\affiliation{%
  \institution{Tsinghua University}
  \country{}}

\author{Guohao Dai}
\affiliation{%
  \institution{Shanghai Jiao Tong University, Infinigence-AI}
  \country{}}


\author{Yu Wang}
\affiliation{%
  \institution{Tsinghua University}
  \country{}}

\thanks{*Corresponding to Yu Wang <yu-wang@tsinghua.edu.cn>, Xiuhong Li <lixiuhong@infini-ai.com>, Guodao Dai <daiguohao@sjtu.edu.cn>}
\renewcommand{\shortauthors}{Hong et al.}
\renewcommand{\shorttitle}{\textit{FlashOverlap}}

\begin{abstract}
  Generative models have achieved remarkable success across various applications, driving the demand for multi-GPU computing. Inter-GPU communication becomes a bottleneck in multi-GPU computing systems, particularly on consumer-grade GPUs. 
  By exploiting concurrent hardware execution, overlapping computation and communication latency becomes an effective technique for mitigating the communication overhead. 
  We identify that an efficient and adaptable overlapping design should satisfy (1) tile-wise overlapping to maximize the overlapping opportunity, (2) interference-free computation to maintain the original computational performance, and (3) communication agnosticism to reduce the development burden against varying communication primitives.
  Nevertheless, current designs fail to simultaneously optimize for all of those features. 
  
  To address the issue, we propose an overlapping design, \hk{named \nickname}, characterized by tile-wise overlapping, interference-free computation, and communication agnosticism. 
  \nickname utilizes a novel signaling mechanism: when part of the output finishes, the computation kernel sends a signal to trigger the communication of that part, while continuing the computation of the remaining part (\textbf{\textit{interference-free computation}}). Consequently, the communication of the finished part and the computation of the remaining part can be overlapped.
  On top of the signaling mechanism, \nickname comprises two key components: (1) the determination of the signaling timing to boost the overlap efficiency (\textbf{\textit{tile-wise overlapping}}), and (2) a pre-communication reordering to create the contiguous address for finished data, enabling communication by simply calling NCCL~\cite{NCCL} APIs (\textbf{\textit{communication agnosticism}}), and a post-communication reordering to correct the data order.
  Experiments show that \nickname achieves up to 1.65$\times$ speedup through overlap, outperforming existing works in most cases. Code is available at \hk{\url{https://github.com/infinigence/FlashOverlap}}. 
\end{abstract}






\maketitle

\section{Introduction}\label{sec:intro}

\begin{figure}[t]
    \centering
    \includegraphics[width=0.98\linewidth]{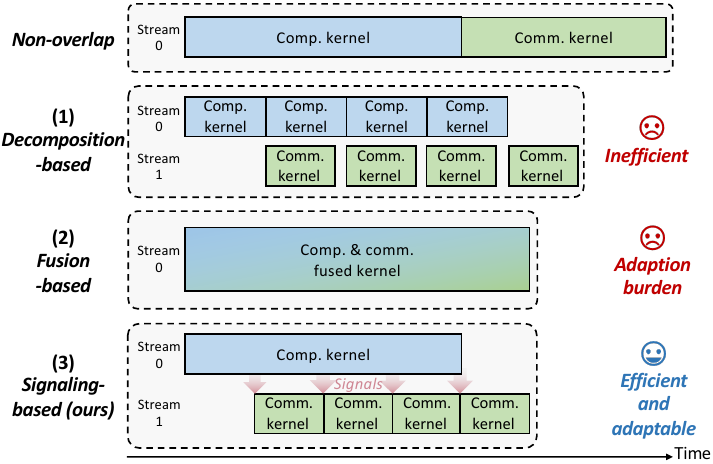}
    \vspace{-0.em}
    \caption{Overlapping methods. (1) Decomposition-based methods are easy to implement while yielding suboptimal overlapping efficiency, (2) fusion-based methods are efficient at the cost of high adaptation efforts, while (3) the proposed signaling-based method optimizes for both efficiency and easy adaptation.}
    \vspace{-0.5em}
    \label{fig:teaser}
\end{figure}

In recent years, generative models have revolutionized various fields, powering applications including chatbots~\cite{deepseekv32024, grattafiori2024llama3herdmodels, chatgpt}, code assistants~\cite{chen2021evaluating, codellama2023}, video generation~\cite{makeavideo2022, hong2022cogvideo, yang2024cogvideox, wan2025}, and agent systems~\cite{liu2023dynamic, huang2024plan}. To continuously enhance the intelligence capabilities of generative models, the number of parameters has dramatically increased, \textit{e.g.}, DeepSeek-V3~\cite{deepseekv32024} contains 671B parameters. Meta has recently previewed its most powerful model, Llama 4 Behemoth~\cite{llama4}, which scales to 2T parameters.
Computing devices, such as GPUs, typically offer limited memory capacity, making it infeasible to accommodate those massive parameters on a single GPU for deploying generative models. Consequently, parameter partitioning across multiple devices has become essential, typically through tensor parallelism (TP), pipeline parallelism (PP), and expert parallelism (EP). Besides, the growing data volume necessitates data parallelism (DP) to enable multi-GPU computation. Such multi-GPU computing paradigms inevitably introduce non-negligible inter-GPU communication overheads, primarily arising from collective communication primitives such as AllReduce, ReduceScatter, and All-to-All.
For deployment on consumer-grade GPUs, such as in local inference scenarios, the communication overhead is further exacerbated, as PCIe interconnection (typically offering 16-64 GB/s bidirectional bandwidth) serves as the primary communication channel between GPUs.

\begin{figure}[t]
    \centering
    \includegraphics[width=0.98\linewidth]{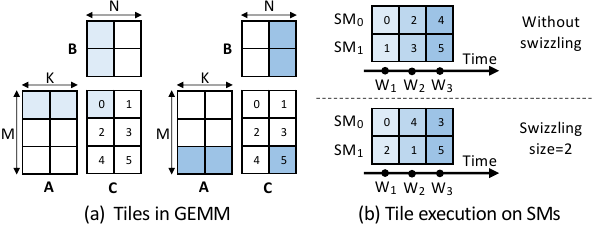}
    \vspace{-0.em}
    \caption{Tile partition and execution in GEMM.} 
    \vspace{-1.0em}
    \label{fig:gemm}
\end{figure}

Overlapping computation and communication has emerged as an effective technique to mitigate communication overhead. The core idea lies in executing computation with communication operations asynchronously, to fully exploit the heterogeneous hardware.
In generative models, the computation part is typically general matrix multiplication (GEMM), and is executed on the high-throughput Tensor Core, while the communication part utilizes specialized interconnection hardware such as NVLink~\cite{NVLink}. 
Unfortunately, data dependency between computation and communication prevents the concurrent execution. 
To resolve the data dependency, two mainstream methods have been proposed, as shown in Fig.~\ref{fig:teaser}. (1) Decomposition-based method decomposes the output tensor of the computation into multiple subtensors, enabling asynchronous overlap between communication for the $k$-th subtensor and computation for the $(k+1)$-th subtensor. The computation and communication of each subtensor form a pair of GEMM and communication kernels, which can be implemented by directly calling APIs such as cuBLAS~\cite{cublasmp2025} and NCCL~\cite{NCCL}, respectively. (2) Fusion-based method fuses the GEMM computation and the communication primitive into a single GPU kernel. In GEMM, a tile refers to a block of output data that is dispatched to streaming multiprocessors (SMs) as a unit for computation, as shown in Fig.~\ref{fig:gemm}. The fusion-based method exploits the inter-tile computation and communication overlap by carefully scheduling their behaviors in the customized kernel.

\begin{table}
  \caption{Comparison of existing works and our design. \hk{\textit{Tile-wise overlapping.} Decomposition-based methods require contiguous data for library API communications, but a 2D GEMM tile is inherently non-contiguous (stride=$N$). \textit{Interference-free computation.} Decomposition-based methods fragment the original GEMM into smaller ones to interleave with communication, and fusion-based methods implement communication into the GEMM kernel. \textit{Communication agnosticism.} Fusion-based methods necessitate adaptation efforts for the customized kernel.}}
  \label{tab:comparison}
  \resizebox{0.48\textwidth}{!}
  {
  \begin{tabular}{lccc}
    \toprule
    \multirow{2}[0]{*}{Method} & Tile-wise & Interference-free & Communication \\
    & Overlapping & Computation & Agnosticism \\
    \midrule
    Decomposition-based & \multirow{2}[0]{*}{$\times$} & \multirow{2}[0]{*}{$\times$} & \multirow{2}[0]{*}{$\bm{\checkmark}$} \\
    \cite{coconetjiangda2022, dominowang2024, compilerwang2023, cetaurichen2024, deepep2025, megascalejiang2024} & & & \\
    \midrule
    Fusion-based & \multirow{2}[0]{*}{$\bm{\checkmark}$} & \multirow{2}[0]{*}{$\times$} & \multirow{2}[0]{*}{$\times$} \\
    \cite{fluxchang2024, fusionamd2024, cometzhang2025, ccfuserwang2025, cublasmp2025, tilelinkzheng2025} & & & \\
    \midrule
    Signaling-based (ours) & {$\bm{\checkmark}$} & {$\bm{\checkmark}$} & {$\bm{\checkmark}$} \\
    \bottomrule
  \end{tabular}
}
\end{table}

However, the two methods fail to jointly optimize for efficiency and adaptability. The inefficiency of the decomposition-based methods arises from two aspects. First, to ensure data contiguity for direct communication API calls, the decomposition is limited to only one dimension, otherwise the subtensor from the decomposition is not contiguous in address. Such a decomposition results in misalignment with the tile-based parallel paradigm for the GEMM, as a tile is inherently non-contiguous in address. Consequently, the decomposition-based method fails to exploit tile-wise fine-grained overlapping.
Second, if the GEMM shape is insufficiently large, the fragmented computation will fail to fully utilize the GPU computational resources, thereby negating the performance benefits of overlap. 
The fusion-based method requires prohibitive manual optimization requirements, which also arise from two aspects.
First, fusion requires manually implemented communication primitives, failing to utilize the existing high-performance communication library such as NCCL~\cite{NCCL}. Consequently, this method suffers from low generalizability: each communication primitive (AllReduce, ReduceScatter, etc.) demands a customized fusion implementation.
Second, when implemented within the same kernel, aligning the data granularity between computation and communication may require modifications to the computational logic or tiling strategies, potentially leading to performance degradation that necessitates further tuning. 

In this paper, we identify that an efficient and adaptable overlapping design should incorporate the following features. (1) Tile-wise overlapping. As tile is logically the minimum parallel data unit in the GEMM output, tile-wise overlapping maximizes the overlapping opportunity. (2) Interference-free computation. To maintain original computational performance, interference with GEMM, including segmentation, tiling or logic changes, should be avoided. (3) Communication agnosticism. The design should be agnostic to the communication primitive, so that no repeated development efforts are spent in implementing the communication primitive.
We categorize the existing works in Table~\ref{tab:comparison}. The decomposition-based methods fail to achieve tile-wise overlapping and interference-free computation, while the fusion-based methods suffer from repeated GEMM tuning and communication primitive implementation efforts. 



We propose a novel signaling-based overlapping design named \nickname that meets all three features. The core idea is utilizing signals to trigger communication without interrupting the GEMM computation process (\textbf{\textit{interference-free computation}}).
On top of that, our design comprises two key components. (1) We analyze and optimize the signaling timing to maximize the overlap efficiency. The signaling mechanism can send a signal for an individual completed tile in GEMM computation (\textbf{\textit{tile-wise overlapping}}), and based on that, we explore the inherent wave pattern in GEMM execution, which means multiple tiles are finished nearly simultaneously as if they are in a wave. Therefore, we unveil the potential of using a wave instead of a tile as the unit of overlapping to enhance the bandwidth utilization with nearly no overlapping opportunity loss. (2) Furthermore, we introduce a pair of reorderings to address the issue caused by data contiguity. Specifically, the pre-communication reordering ensures the contiguous address of data ready for communication, which enables the direct NCCL~\cite{NCCL} API calls for implementation (\textbf{\textit{communication agnosticism}}). The post-communication reordering is carefully designed to correct the data order after the communication.
To further enhance the performance of our design, we extend the signaling timing to be a tunable number of waves (denoted as a wave group in this paper), and further propose a predictive search method to find the optimal wave grouping solution in real time.

In summary, this paper makes the following contributions:
\begin{itemize}
    \item We introduce a novel signaling-based design to achieve computation-communication overlap, which sends signals from the GEMM kernel to trigger communication without interrupting the computation process. 
    \item Starting from the tile-wise signaling, we exploit the wave-wise signaling to enhance bandwidth utilization while maintaining the overlapping opportunity. We further enable the design to be tunable via wave grouping, and propose a predictive search method to optimize the wave grouping selection. 
    \item We introduce a pair of reorderings before and after communication, with the former creating a contiguous data address for NCCL~\cite{NCCL} API calling, and the latter correcting the data order. The overhead of reorderings is mitigated by kernel fusion.  
    \item We conduct experiments to evaluate the proposed design with communication primitives including AllReduce, ReduceScatter, and All-to-All, with each tested under hundreds of GEMM sizes. We also evaluate the design in the inference and training tasks with typical generative models. The results demonstrate that our design achieves up to 1.65$\times$ speedup through computation-communication overlap, and is effective in end-to-end inference and training. 
\end{itemize}

\section{Background}\label{sec:background}
In this section, we elaborate on the characteristics of the GEMM computation and the existing inter-GPU communication implementations on modern GPUs. Subsequently, we demonstrate that the pattern of GEMM computation followed by data-dependent communication can be commonly found in both training and inference of generative models, emerging as one of the primary bottlenecks for improving efficiency in multi-GPU computing systems. Based on that, we present a comprehensive survey and comparative analysis of prior works for computation-communication overlap.

\subsection{General Matrix Multiplication}

\subsubsection{Wave Pattern in GEMM} 
As the core operator in neural networks, general matrix multiplication (GEMM) can be formulated as $A^{M\times K} \times B^{K\times N} = C^{M \times N}$, where $M, N, K$ collaboratively represent the GEMM size. Modern GPUs consist of multiple streaming multiprocessors (SMs)~\cite{gpusm}, where each SM contains independent computational and on-chip memory resources. To exploit the parallel execution across SMs, a GEMM workload is partitioned into tiles distributed across SMs. The output matrix $C$ is partitioned into tiles, with each tile's workload including the corresponding data loading and computation from the input matrices $A$ and $B$. Those tiles are scheduled across different SMs for parallel execution.  A concrete example is illustrated in Fig.~\ref{fig:gemm}, where six tiles are distributed across two SMs. Consequently, the tile execution follows a specific sequential order. Notably, the completion time of the tiles exhibits a distinct wave pattern.

\begin{figure}[t]
    \centering
    \includegraphics[width=0.98\linewidth]{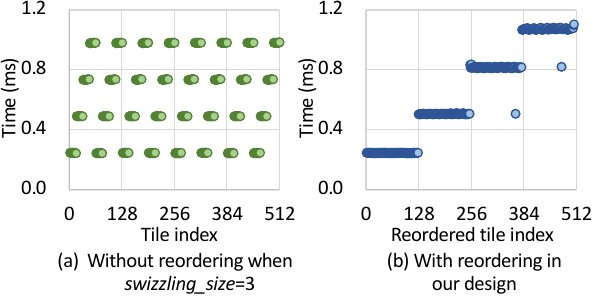}
    \vspace{-0.em}
    \caption{Wave pattern in GEMM execution. Each point in (a) and (b) represents the corresponding completion time of each tile, and the time is captured by the global timer~\cite{globaltimer}.} 
    \vspace{-1.0em}
    \label{fig:wave}
\end{figure}

A wave is defined as a set of concurrently executed tiles~\cite{osama2023streamk}. As shown in Fig.~\ref{fig:wave}, we record the completion time of each tile in a GEMM ($M=2048, N=K=8192$) on an RTX 4090 GPU, and the tile completion time can be distinctly categorized into four distinct waves, which is consistent with the result of dividing tile number (512) by SM number (128). Furthermore, we observe that the completion order of tiles does not align with the memory address (represented by tile index), \hk{if the block swizzling is applied, as detailed in Sec.\ref{sec:swizzle}}. 

\subsubsection{Block swizzling}\label{sec:swizzle}
\hk{The tile execution order in GEMM is influenced by techniques such as block swizzling~\cite{CUTLASS}. Block swizzling refers to scheduling tiles onto SMs in a swizzling manner for enhanced memory access efficiency, as depicted in Fig.~\ref{fig:gemm}(b). The address discontiguity in a wave prevents early-finished tiles from being promptly communicated. To address the mismatch and enable tile-wise overlapping, we introduce the data reordering technique, which is described in Sec.~\ref{sec:map}.}


\subsubsection{Main Loop and Epilogue}
The execution of a GEMM involves the main loop and the epilogue. The main loop performs the core multiply-accumulate operations and accounts for the majority of the GEMM duration, while the epilogue refers to element-wise operation (\textit{e.g.}, ReLU, SiLU, or bias addition) performed after matrix multiplication. Those element-wise operation is typically fused with the preceding matrix multiplication into a single GPU kernel~\cite{evtchen2024}, thereby eliminating redundant memory accesses and kernel launch overhead.

\subsection{Inter-GPU Communication}
The underlying behaviours of inter-GPU communication vary significantly across different hardware configurations (intra-node or inter-node, via NVLink~\cite{NVLink}, PCIe~\cite{PCIe}, or InfiniBand~\cite{infiniband}, etc.). Aimed at hiding the interconnect hardware complexities from users, libraries such as the NVIDIA collective communications library (NCCL)~\cite{NCCL} provide high-level APIs for various communication primitives. NCCL supports encapsulated collective communication primitives including AllReduce, AllGather, and ReduceScatter, as well as point-to-point send/receive operations, which can be used for constructing the All-to-All primitive. Besides the convenience, NCCL also optimizes efficiency for communication, such as implementing the RING algorithm~\cite{patarasuk2009ring} for better bandwidth efficiency. The communication agnostism enables our design to leverage NCCL through standard API calls directly. Note that other libraries such as MSCCLang~\cite{cowan2023mscclang} and DeepEP~\cite{deepep2025} are achieving similar functionality, and can also be seamlessly integrated into our design through API calls. 

\begin{figure}[t]
    \centering
    \includegraphics[width=0.98\linewidth]{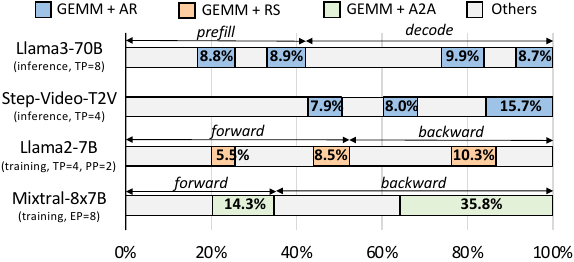}
    \vspace{-0.em}
    \caption{Typical time portion of "GEMM + X" in inference and training. All profilings are on A800 GPUs.} 
    \vspace{-1.0em}
    \label{fig:timeline}
\end{figure}


\subsection{Computation and Data-Dependent Communication}

In multi-GPU computing systems, each GPU is responsible for only part of the computation, thereby necessitating communication for data exchange and synchronization. Such a pattern is highly prevalent, as illustrated as follows:

\subsubsection{GEMM followed by AllReduce (GEMM+AR)} Applying parallelism methods to model inference or training frequently triggers an AllReduce primitive after a GEMM. Specifically, the tensor parallelism (TP) needs an AllReduce operation to reduce the GEMM partial results from all the GPUs in the parallel group~\cite{shoeybi2019megatronmp}, and the data parallelism (DP) uses an AllReduce operation to compute gradient summation across all GPUs~\cite{li2020torchdp}. GEMM+AR is widely utilized, but it incurs significant communication overhead due to the high complexity of the AllReduce primitive. We test typical scenarios with TP including LLM serving and text-to-video (T2V) generation, and the latency breakdown in Fig.~\ref{fig:timeline} shows GEMM+AR occupies 31.6\%-42.2\% of the end-to-end duration. 
\subsubsection{GEMM followed by ReduceScatter (GEMM+RS)} In model training with TP, AllReduce is typically decomposed into a ReduceScatter and an AllGather, and the ReduceScatter and its preceding GEMM form a GEMM+RS pattern. Besides, the backward pass of fully shared data parallelism (FSDP) performs ReduceScatter on weight gradients after GEMM computation~\cite{zhao2023fsdp}. As shown in Fig.~\ref{fig:timeline}, GEMM+RS takes roughly 30\% of the end-to-end time in training a Llama2-7B model.  
\subsubsection{GEMM followed by All-to-All (GEMM+A2A)} The widely adopted Mixture-of-Experts (MoE) models~\cite{lepikhin2021gshard} typically employ expert parallelism (EP) to distribute experts across multiple GPUs. In this paradigm, as data is dynamically routed to specific experts on particular GPUs, an All-to-All communication operation is necessary to transfer the processed data back to the original GPUs after expert computation. Specifically, the MoE part contains linear layers, which leads to the GEMM+A2A pattern. Notably, the dynamic routing mechanism creates inherent workload imbalance among GPUs, exacerbating the communication overhead. To quantify such an overhead, we profile the training of a Mixtral-8x7B model and observe that GEMM+A2A achieves over 40\% of the overall latency. 

\subsection{Related Works}

Prior research has extensively investigated the overlapping technique to mitigate communication bottlenecks in multi-GPU computing systems. Regarding overlapping techniques, existing works can broadly be classified into two categories:  (1) overlapping data-dependent computation and communication, and (2) leveraging existing multiple dataflows to enable overlap. For the first category concerning data-dependent computation and communication, decomposition-based and fusion-based methods are the predominant approaches. 

\subsubsection{Decomposition-based Method}
In CoCoNet~\cite{coconetjiangda2022}, the authors point out the importance of scheduling computation order to align with the communication order, and further design a compiler-based approach to automatically generate efficient GPU kernels that coordinate computation and communication. \cite{compilerwang2023} also utilizes a compiler-based approach, introducing a cost model to handle the trade-offs between overlapping opportunity and communication segmentation. 
Domino~\cite{dominowang2024}, Async-TP~\cite{asynctp}, and MegaScale~\cite{megascalejiang2024} have applied the decomposition-based method to LLM training in practice.
Centauri~\cite{cetaurichen2024} builds a comprehensive communication partition space and performs hierarchical scheduling to maximize overlapping efficiency in LLM training. While careful optimization of decomposition strategies (\textit{e.g.}, decomposition granularity and dimension) enhance performance, decomposition-based methods fundamentally cannot achieve tile-wise overlapping, and hence the potential improvement remains limited.

\subsubsection{Fusion-based Method}
In \cite{fusionamd2024}, the AMD team designs the fusion paradigms of embedding+All-to-All, general matrix-vector multiplication (GEMV)+AllReduce, and GEMM+All-to-All on AMD GPUs. FLUX~\cite{fluxchang2024} optimizes for TP, fusing the communication primitive into the beginning or end of the highly optimized GEMM kernel at the tile level, by sharing the address used in remote access. Comet~\cite{cometzhang2025} introduces a thread block specialized kernel to implement the computation-communication overlap for MoE layers, distributing computation and communication to different SMs for parallel execution, where the SM ratio between the two operations can be adjusted for better efficiency. To reduce the development effort, TileLink~\cite{tilelinkzheng2025} introduces a compiler-based approach to automatically generate the overlapping GPU kernel using tile-centric primitives. NVIDIA also develops cuBLASMp~\cite{cublasmp2025} to support such fusion. Although implementation details differ, fusion-based methods universally adopt tile-wise overlapping to achieve improved overlapping efficiency. However, kernel fusion creates new demands for communication implementation and optimization, and necessitates logic or tiling changes when coordinating computation and communication pipelines. Besides, T3~\cite{pati2024t3} explores fine-grained fusion under non-invasive GEMM modifications, tracking the progress of tiles to trigger communication. However, T3 relies on a specific hardware design and is evaluated in simulation, which still faces challenges in real systems. 

\subsubsection{Multi-dataflow Scheduling}
Another line of research~\cite{dhelixwang2024, deepseekv32024, lancetjiang2024, fastermoehe2022, zhu2024nanoflow} exploits multi-dataflow scheduling to achieve overlap. Those works exploit inherent parallel data dependencies (\textit{i.e.}, multi-dataflows), including those between forward and backward passes, among MoE experts, and between weight and activation gradients during backpropagation, to achieve computation and communication overlap across different dataflows. \hk{Compared to multi-dataflow scheduling methods, the decomposition-based and fusion-based methods target data-dependent overlap between computation and communication, involving only the communication operator and its adjacent computation operator, which is also the focus of this work. \nickname maintains orthogonal compatibility with multi-dataflow scheduling methods to expand design space, potentially enabling computation-communication overlap within individual dataflows.}

\begin{figure*}[t]
    \centering
    \includegraphics[width=0.98\linewidth]{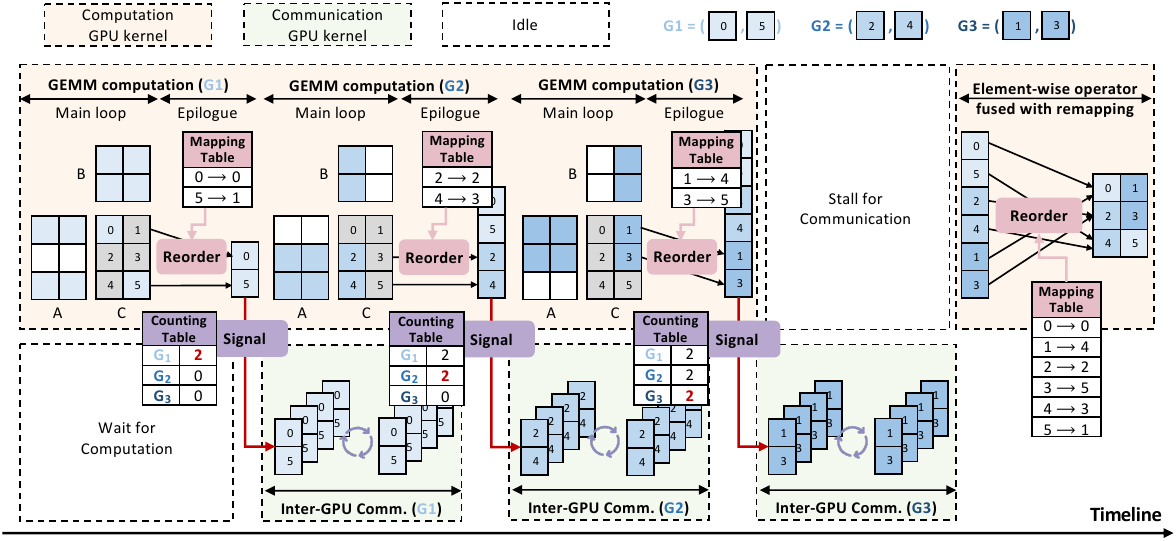}
    \vspace{-0.5em}
    \caption{System overview. The GEMM computation is executed within one GPU kernel, and when each group ($G_1$, $G_2$, or $G_3$) of tiles finishes, it first reorders the tiles in the group to contiguous addresses, and then signals to trigger the corresponding inter-GPU communication of the group. To correct the order, the tiles are reordered back when communication finishes.} 
    \vspace{-0.em}
    \label{fig:framework}
\end{figure*}

\begin{table}
  \vspace{.0em}
  \caption{Notation description.}\label{tab:symbol}
  \resizebox{0.4\textwidth}{!}{
  \begin{tabular}{l|l}
    \toprule
    Notation & Description \\
    \midrule
    $M$ & Input dimension in a GEMM \\
    $N$ & Output dimension in a GEMM \\
    $K$ & Accumulation dimension in a GEMM \\
    $T$ & Number of waves \\ 
    $P$ & Number of wave groups \\
    $W_i$ & The $i$-th wave \\
    $G_j$ & The $j$-th wave group \\
    \bottomrule
  \end{tabular}
  }
  \vspace{-0.5em}
\end{table}

\section{System Design}\label{sec:system}
In this section, we introduce the system design of \nickname. Following the overview, we present detailed descriptions of the two core technologies: signaling and reordering. First, we revisit the design motivation behind each technique, then systematically clarify its respective challenge, insight, and approach. In the end, we describe the design space for performance tuning.

\subsection{Overview}
Fig.~\ref{fig:framework} illustrates the overview of the proposed design. The GEMM computation remains a single GPU kernel with the proposed overlapping design, sending signals to trigger communication. Considering data dependency, the signal ensures that the communicated tiles finish the corresponding GEMM computation. Moreover, a pair of reordering operations is introduced to create contiguous addresses for communication and to correct the data order.

\subsection{Signaling}\label{sec:signal}
\subsubsection{\textbf{Motivation: Data Dependency.}} 
The signaling mechanism is proposed to track the fully computed data that is ready for communication without bringing interference to the GEMM computation. 
In decomposition-based methods~\cite{coconetjiangda2022, dominowang2024, compilerwang2023, cetaurichen2024, deepep2025, megascalejiang2024}, the communication of a subtensor is directly triggered upon completion of the corresponding GEMM computation. Fusion-based methods~\cite{fluxchang2024, fusionamd2024, cometzhang2025, ccfuserwang2025, cublasmp2025, tilelinkzheng2025} employ instruction scheduling to chain the dependent computation and communication operations. However, both methods bring interference to the native GEMM computation, as demonstrated in Sec.~\ref{sec:intro}. To avoid such interference, ideally, when a certain part of the data finishes GEMM computation, a signal with negligible overhead is used to initiate the corresponding communication while the GEMM kernel continues the computation.  In this way, the signal chains the data dependency without computation interference and handcrafted communication implementation. 

\subsubsection{\textbf{Challenge: Signaling Timing.}}
It is challenging to determine the timing of signaling for communication, as a higher opportunity of overlapping does not equal a higher speedup. 
As mentioned in Sec.~\ref{sec:intro}, a tile is the minimal parallel data unit in GEMM output and brings the maximum overlapping opportunity. 
However, directly triggering communication for each finished tile leads to significant communication fragmentation. The latency of tile-by-tile communication can be excessive, emerging as the bottleneck that invalidates the overlap.
Fig.~\ref{fig:curve} depicts the bandwidth curves varying with the data amount on RTX 4090 and A100 GPUs, and we discover a sharp degradation when the data amount falls below a certain threshold. Taking the AllReduce primitive on four RTX 4090 GPUs as an example, the communication of a tile (192KB) yields only 13\% of the bandwidth utilization.
Therefore, considering the trade-off between overlapping opportunities and communication segmentation, the signaling timing necessitates further optimization. 


\subsubsection{\textbf{Insight: from Tile to Wave, from Wave to Group. }}
Instead of signaling for a tile, the proposed mechanism signals for a wave of tiles together. As detailed in Sec.~\ref{sec:background}, we investigate the wave pattern in GEMMs. The wave pattern denotes that certain tiles (\textit{i.e.,} a wave) are completed nearly simultaneously, typically within 5\% of a wave duration. Therefore, the tile-wise signaling is not necessary, and we can directly use a wave of tiles for signaling to trigger the communication, which yields essentially the same overlapping opportunity and better bandwidth utilization. 

Furthermore, we observe that static wave-wise signaling is not optimal against workload variety. A trade-off exists between smaller but immediate communication and larger but delayed communication. Thus, we design the signaling timing to be tunable and define the wave group on top of the waves. A group $G$ includes $|G| \geq 1$ waves, and the corresponding communication starts after each group finishes the GEMM computation. The size of each group is tunable. 

\begin{figure}[t]
    \centering
    \includegraphics[width=0.98\linewidth]{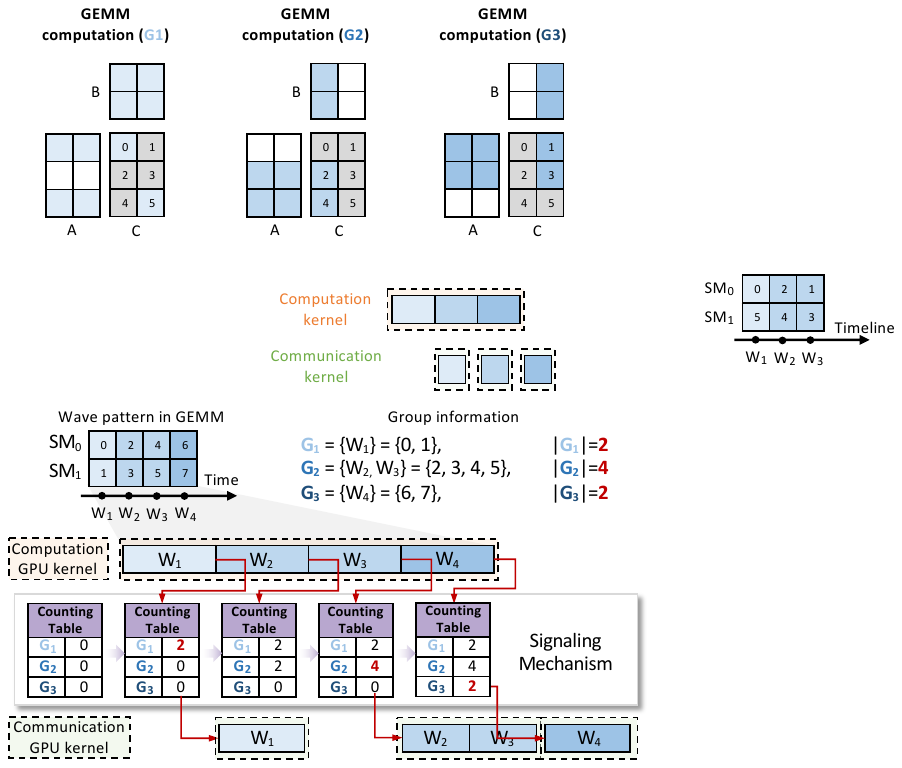}
    \vspace{-0.em}
    \caption{Group-wise tile counting for signaling.} 
    \vspace{-1.0em}
    \label{fig:signal}
\end{figure}

\begin{figure*}[t]
    \centering
    \includegraphics[width=0.98\linewidth]{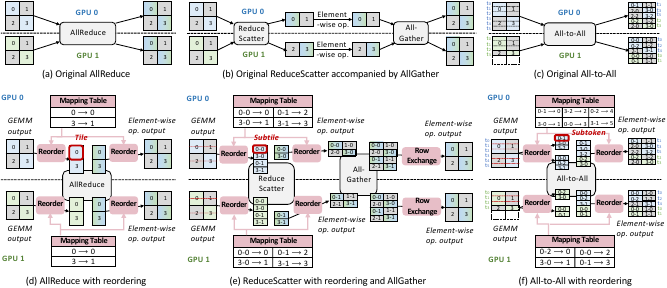}
    \vspace{-0.em}
    \caption{Pre-communication and post-communication reordering patterns under different communication primitives.} 
    \vspace{-0.5em}
    \label{fig:mapping}
\end{figure*}

\subsubsection{\textbf{Approach: Group-wise Tile Counting. }} 
We introduce a counting table to track the completion of tiles, and the finished tiles are recorded separately by groups. Specifically, the counting table is of size $P$, indicating the tiles are divided into $P$ different groups ($G_1, G_2, ..., G_P$). The $j$-th number in the counting table is atomically added by 1 when a tile in $G_j (j \leq P)$ is finished. Once the $j$-th number reaches $|G_j|$, the communication of $G_j$ starts. We simply use the tile index to identify which group a tile belongs to. 

As shown in Fig.~\ref{fig:signal}, there are 3 groups ($G_1, G_2, G_3$), with each including tiles of 1 or 2 waves. The counting table is initialized to all zeros. When the first wave $W_1$ finishes, tile 0 and 1 are recorded for $G_1$ in the table. The counting number reaches $G_1=2$, thereby signaling the communication of the tiles in $G_1$ (also $W_1$) to start. For $G_2$ with 2 waves (4 tiles), the communication holds until the counting number reaches 4, and all four tiles are communicated together. The behavior of $G_3$ is the same as that of $G_1$.

\subsection{Reordering}\label{sec:map}
\subsubsection{\textbf{Motivation: Contiguous Address for Communication.}}
A single inter-GPU communication behavior through calling the NCCL~\cite{NCCL} library necessitates contiguous addresses for both the sending and receiving buffers. 
To enable flexible communication of tiles, both the intra-tile and inter-tile data that are communicated together should be reordered into a contiguous address. 

\subsubsection{\textbf{Challenge: Irregular Tile Execution Order.}} 
Besides the intra-tile data being inherently non-contiguous, the inter-tile execution order tends to be irregular in GEMMs. The underlying reason for such irregularity is the application of block swizzling~\cite{CUTLASS}, which is an optimized technique for GEMM performance. Block swizzling schedules the tiles onto GPU blocks in a swizzling manner for enhanced memory access efficiency. Fig.~\ref{fig:gemm} (b) shows a typical tile execution order with swizzling size $=2$. After the first wave $W_1$, the finished tiles 0 and 2 do not form a contiguous data block. Therefore, the challenge lies in achieving both intra-tile and inter-tile data contiguity within a wave group through reordering, while ensuring communication correctness.  

\subsubsection{\textbf{Insight: Data Order Can be Incorrect. }}\label{sec:mapping_insight}
The correct data order is not strictly required for inter-GPU communication, which allows us to reorder those incontiguous tiles to form a large and contiguous data block. (1) AllReduce. The requirement is to maintain consistent tile order across all GPUs, while this tile order can be entirely different from that in the original GEMM output matrix without affecting the correctness of communication. (2) ReduceScatter. The tensor is reduced and uniformly sliced along the row dimension to be distributed onto different GPUs. 
We emphasize that the GPU assignment for each row is not essential, as ReduceScatter is typically paired with AllGather, and all the rows are aggregated by the subsequent AllGather. However, it is necessary to guarantee that each row resides entirely on a specific GPU, so that it can properly compute element-wise operations (\textit{e.g.}, normalization) before AllGather. Based on that, the core idea is to allow mismatched row ordering to gain more overlapping opportunities. Assuming tile 0 and tile 1 in Fig.~\ref{fig:mapping} (b) are in the first wave, under the ReduceScatter primitive, they are both assigned to GPU 0. Such a situation prevents the early ReduceScatter communication on tiles 0 and 1, leading to a shrunk overlap opportunity. (3) All-to-All. The data division across different GPUs is at the token (\textit{i.e.}, row) level, and a row is specifically dedicated to a determined GPU. To conclude, although communication primitives bring some limitations on the data order for communication correctness, there is still space remaining for reordering. 

\subsubsection{\textbf{Approach: Execution-Order-Aware Reordering}}
As shown in Fig.~\ref{fig:mapping} (d)-(f), to ensure intra-tile data contiguity, the reordered output is first reshaped into a column of tiles (row major) before communication. 
For the inter-tile data contiguity, the tile indices are reordered based on their execution order, and we introduce a mapping table to record the index mapping. 

(1) AllReduce. The tiles in the early finished wave are reordered to be in the front, and the tiles within the same wave are reordered to be together. The relative index order of tiles within the same wave can be random, as those tiles are always communicated together. Fig.~\ref{fig:mapping} (d) shows an example where the reordered indices of tile 0 and tile 3 are 0 and 1, respectively. Such an indexing method is generally useful for all primitives. Through a pair of reorderings, the output in Fig.~\ref{fig:mapping} (d) is the same as that in Fig.~\ref{fig:mapping} (a). 

(2) ReduceScatter. To ensure a row is complete on a specific GPU, each tile is first split equally across the row dimension to form subtiles as many as GPU number. Instead of a tile, a subtile is utilized as the reordering unit. In this way, no matter how the tiles are assigned to GPUs, the $k$-th subtile within a tile always resides on the $k$-th GPU at the end, and all of those $k$-th subtiles form complete rows. Note that the mapping table should be adjusted accordingly, as shown in Fig.~\ref{fig:mapping} (b). Although such reordering leads to row assignment changes compared to the original ReduceScatter in Fig.~\ref{fig:mapping} (b), the subsequent AllGather aggregates the rows together, and the row order can be corrected by a local row exchange. The row exchange needs no mapping table, as it is simply a block cyclic permutation, which can also be seamlessly fused into element-wise kernels. Alternatively, if the row order has no impact on subsequent processing, the exchange step can be safely eliminated. In this way, the element-wise operation before AllGather is properly computed, as each row is complete on a single GPU.

(3) All-to-All. Each tile is split by the token (\textit{i.e.}, row) to form subtokens. We introduce a specific memory pool for each destination GPU to exclusively store the corresponding subtokens, so that the subtokens are communicated to their destination GPUs and form the complete tokens at the end. Within each memory pool, the subtokens are still reordered based on the execution order. When a wave group finishes computation and sends the signal, the communication for all the subtokens in all the memory pools starts, with the subtokens in each pool being sent to the destination GPU. The two reorderings ensure the output in Fig.~\ref{fig:mapping} (c) is the same as that in Fig.~\ref{fig:mapping} (f), where $t_i$ denotes the $i$-th token. 

The pre-communication reordering is fused into the GEMM, involving only the epilogue without interrupting the main loop. After the communication, we need a reordering operation to restore the original data order, which can be seamlessly fused into the subsequent element-wise kernel. Specifically, instead of directly storing/loading data (tile, subtile, or subtoken) based on the original index, the fused kernel stores/loads data based on the reordered index. Since the mapping table's size is negligible compared to the input matrix, it brings nearly no extra memory access.  Although the access pattern is changed, the memory efficiency is well preserved due to the guaranteed locality of sufficiently contiguous data. We provide a comprehensive quantitative analysis of the associated overhead in Sec.~\ref{sec:eval_overhead}.

\begin{figure}[t]
    \centering
    \includegraphics[width=0.98\linewidth]{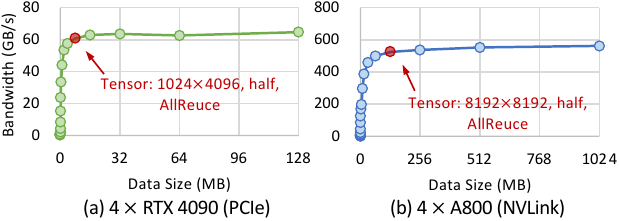}
    \vspace{-0.em}
    \caption{Bandwidth curve varying with data size. The red spots are the borderline facing bandwidth degradation. } 
    \vspace{-0.5em}
    \label{fig:curve}
\end{figure}

\begin{figure}[t]
    \centering
    \includegraphics[width=0.98\linewidth]{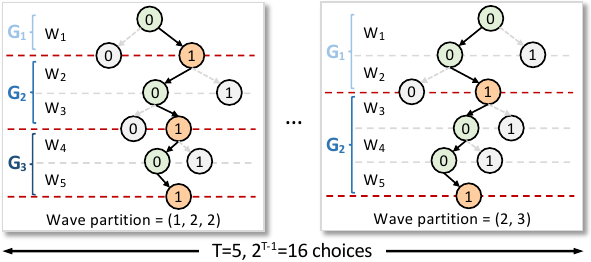}
    \vspace{-0.em}
    \caption{Design space of wave group partitions.}
    \vspace{-0.5em}
    \label{fig:design_space}
\end{figure}

\subsection{Design Space}\label{sec:design_space}
The design is characterized by the tunable configuration of wave grouping to optimize the overlapping performance. We formulate a binary discrete decision optimization problem, and the optimization objective is to minimize the latency after overlap. After each wave, the tiles can be decided to be communicated (denoted as "1") or not (denoted as "0"). The last wave is the exception, as all the accumulated tiles must be communicated. Assuming there are $T$ waves, the design space is of size $2^{T-1}$. Consider the first example in Fig.~\ref{fig:design_space}, we choose to conduct communication after $W_1, W_3$ and $W_5$, thus deriving the wave group partition of $(1, 2, 2)$. The group sizes are $|G_1| = 1, |G_2| = 2, |G_3| = 2$, respectively. In the second example, the communication is triggered after $W_2$ and $W_5$, with the wave group partition being $(2, 3)$. Such a partition leads to two groups with sizes of 2 and 3, respectively.


\section{Real-Time Tuning}
In this section, we first introduce the predictive search method applied in real-time searching for the wave group partition, then describe the whole tuning algorithm for our design.

\subsection{Predictive Search}\label{sec:search}
We introduce a predictive search method for the tunable wave grouping. 
The proposed method reduces the design space based on some prior knowledge, and then replaces the online profiling with a latency predictor, which enables the generation of the optimal wave group partition in real time. 

\subsubsection{\textbf{Motivation: Tuning is Necessary. }}
 Tuning the wave group partition is essential for the performance. We take a group containing only a wave as the baseline partition. Theoretically, the baseline partition achieves the most fine-grained overlapping. However, we conduct experiments and discover that it fails to deliver the optimal performance in most cases. Specifically, among over 50 GEMM sizes tested with the AllReduce primitive on four RTX 4090 GPUs, only 4\% point to the baseline partition under exhaustive search. Employing the baseline partition leads to an average of 17.34\% performance degradation. Fundamentally, such degradation results from the under-utilized bandwidth due to segmented communication, and the overhead of frequent API calls, which collectively become the performance bottleneck when communication latency dominates.  

\subsubsection{\textbf{Challenge: Tuning Overhead. }}
The original tuning design space is of size $2^{T-1}$, with each partition candidate requiring an online execution to select the optimal, thus leading to non-negligible tuning overhead. A typical GEMM output size $M = 4096, N = 8192, K = 7168$ in computing generative models yields 8 waves ($T$=8) when computed on an NVIDIA RTX 4090 GPU, equaling a design space of 128 candidates. The online execution takes approximately 5 ms, and the profiling typically includes 10 warm-up and 100 timing tests to mitigate the measurement fluctuations. Therefore, the online profiling takes more than 1 minute (>100$\times$ of model forwarding latency), which is unacceptable for end-to-end performance. 

\subsubsection{\textbf{Insight: Overlap Latency Analysis. }}
The design space can be reduced based on prior knowledge, and the online profiling can be replaced by a predictor that predicts the latency. Considering the composition of the latency with overlap, the overlap part occupies the middle part of the timeline, and hence, the head and tail parts of the timeline are influential. The head and the tail are determined by the first and the last group sizes, respectively. Thus, both the first and the last group sizes are preferred not to be too large, to avoid the cold start and the long tail. The design space can be reduced by leveraging such a principle.

Furthermore, instead of online profiling, we can design an accurate latency predictor for the searching process. To achieve high prediction accuracy, the separate latency of each operation after overlap needs analysis. (1) Computation. Since the main loop in GEMM is well preserved, the latency is mainly affected by the computational resource contention. (2) Communication. The influential factor is the communication segmentation, leading to a prolonged latency. Based on that, the wave group partition further decides the overlap pattern. 

\subsubsection{\textbf{Approach: Design Space Pruning and Predictive Search. }}
The proposed search method constrains the sizes of the first and last groups to prune the design space, and applies a latency predictor in the searching process. Specifically, setting $|G_1| \leq S_1$ and $|G_{P}| \leq S_P$, the design space size reduced to $\mathcal{O}(2^{T-2})$. We use $S_1=2$ and $S_P=4$ for evaluation. 

Since our design sets a higher priority for the communication kernel, once the SM number occupied by the communication primitive is determined, the remaining SMs are available for the GEMM computation. Considering that the utilized SMs become fewer, the latency of GEMM computation after overlap is derived by adjusting the original latency according to the updated wave number. We estimate the communication latency based on the bandwidth curve that varies with data amount, which is depicted in Fig.~\ref{fig:curve}. Specifically, the communication latency of each group is predicted with its containing data amount as the input, and the total communication latency is the sum of all. The detailed predicting method is illustrated in Alg.~\ref{alg:tuning}.

\begin{algorithm}[t]
    \caption{Grouping tuning algorithm} 
    \label{alg:tuning}
    \begin{flushleft}
    \textbf{Input:} $M$, $N$, $K$ \textcolor{blue}{\# GEMM size}, \textit{comm\_op} \textcolor{blue}{\# communication primitive}, \textit{gpu} \textcolor{blue}{\# GPU hardware}
    \end{flushleft}
    \begin{algorithmic}[1]
        \State \textcolor{blue}{\# Offline: get GEMM configuration}
        \State \textit{gemm\_config} $\gets$ get\_config($M$, $N$, $K$, \textit{gpu})
        \State $T$ $\gets$ \textit{gemm\_config.tile\_num} / (\textit{gpu.sm\_num} - \textit{comm\_op.sm\_num})
        \State \textcolor{blue}{\# Offline: get the (data size, bandwidth) curve}
        \State \textit{bdw\_curve} $\gets$ sample\_bandwidth(\textit{comm\_op}, \textit{gpu})
        \State \textcolor{blue}{\# Online: tuning wave grouping partition}
        \State \textit{candidates} $\gets$ get\_candidiates($T$)
        \State $t^{\min} \gets +\infty$
        \For{$\bm{G}$ \textbf{in} \textit{candidates}}
        \State $t^{\text{acc}}_{p}$ $\gets$ 0, $t^{\text{acc}}_{m}$ $\gets$ 0
        \For{$i, G_i$ \textbf{in} enumerate ($\bm{G}$)}
        \State \textcolor{blue}{\# Interpolate the comm. latency of the last group}
        \State \textit{data\_size} $\gets$ get\_data\_size($G_{i-1}$)
        \State $t_{m}$ $\gets$ interp\_latency(\textit{bdw\_curve}, \textit{data\_size})
        \State \textcolor{blue}{\# Calculate computation latency of this group}
        \State $t_{p}$ $\gets$ \textit{gemm\_config.duration}$/ T \times |G_i|$
        \State \textcolor{blue}{\# Latency accumulation}
        \State $t^{\text{acc}}_{m}$ $\gets$ $\max$($t^{\text{acc}}_{p}$, $t^{\text{acc}}_{m}$)  + $t_{m}$, $t^{\text{acc}}_{p}$ $\gets$ $t^{\text{acc}}_{p}$ + $t_{p}$
        \EndFor
        \State \textcolor{blue}{\# Add the communication latency of the final group}
        \State \textit{data\_size} $\gets$ get\_data\_size($G_{-1}$)
        \State $t^{\text{acc}}_{m}$$\gets$ $\max$($t^{\text{acc}}_{p}$, $t^{\text{acc}}_{m}$) + interp\_latency(\textit{bdw\_curve}, \textit{data\_size})
        \State \textcolor{blue}{\# Get the optimal wave partition}
        \If {$t^{\text{acc}}_{m}$ < $t^{\min}$}
        \State $t^{\min}$ $\gets$ $t^{\text{acc}}_{m}$, $\bm{G}^{\text{optimal}}$ $\gets$ $\bm{G}$
        \EndIf
        \EndFor
        \State return $\bm{G}^{\text{optimal}}$
    \end{algorithmic}
    \vspace{-0.em}
\end{algorithm}

\subsection{Tuning Algorithm}
As shown in Alg.~\ref{alg:tuning}, the tuning algorithm can be divided into the offline and online stages. \hk{After deployment, the model} \hk{architecture, hardware, and network topology are fixed, and the tuning is repeated only for new GEMM sizes. Thus, we define the offline stage as the procedure of handling the deployment setups, and the online stage as the repeated tuning for different GEMM sizes. } The offline stage is responsible for deriving the GEMM configuration, the communication bandwidth curve, and figuring out the resource contention on SMs. At the online stage, we search for the optimal partition in the design space based on latency prediction. 

\subsubsection{Offline Stage}
(1) Computation. Given a problem size $M\times N\times K$, GEMM configurations are typically available, leveraging existing highly optimized linear algebra implementations (\textit{e.g.}, cuBLAS~\cite{cuBLAS} and CUTLASS~\cite{CUTLASS} on NVIDIA GPUs). The required GEMM configurations include tiling size, sizzling pattern, and the corresponding duration, etc. (2) Communication. Performing a communication primitive on given GPUs, the bandwidth exhibits continuous variation with the data size, as shown in Fig.~\ref{fig:curve}. Therefore, the bandwidth curve is sampled with multiple dense points, given a data size, and the effective bandwidth can be accurately estimated through interpolation of sampled points. (3) Resource contention. On given GPUs, the SM number occupied by a communication primitive using NCCL~\cite{NCCL} can be derived. Thus, the total wave number of the GEMM computation is updated according to line 3 in Alg.~\ref{alg:tuning}.

\begin{figure*}[t]
    \centering
    \includegraphics[width=0.98\linewidth]{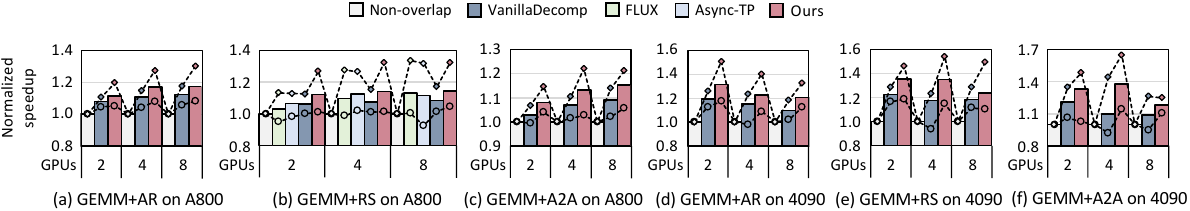}
    \vspace{-0.em}
    \caption{Operator-level average speedup on average. "$\circ$" and "$\diamond$" denote the minimum and maximum speedups, respectively.} 
    \vspace{-0.em}
    \label{fig:opertor}
\end{figure*}

\begin{figure*}[t]
    \centering
    \includegraphics[width=0.98\linewidth]{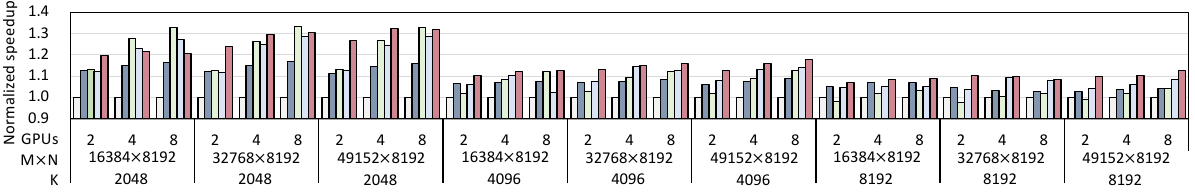}
    \vspace{-0.em}
    \caption{Operator-level speedup comparison on typical shapes, tested with GEMM+RS on A800 GPUs.} 
    \vspace{-0.em}
    \label{fig:op-a800}
\end{figure*}

\subsubsection{Online Stage}
(1) Generate the candidates (line 7). The original design space is the binary decision at each wave, forming a space size of $2^{T-1}$. The algorithm reduces the design space with prior knowledge by constraining group size, as mentioned in Sec.~\ref{sec:search}. The candidate partitions are stored in \textit{candidates}. (2) Predictive search (lines 10-22). All candidate partitions are compared based on their predicted latencies. The predicted latency is accumulated by looping over the groups in the corresponding partition. Specifically, we interpolate the last group's communication latency ($t_{m}$) and derive the current group's computation latency ($t_{p}$) based on its wave number. Based on that, the proposed algorithm accumulates the communication and computation latencies separately. Specifically, the accumulated computation latency ($t_{p}^{\text{acc}}$) is accumulated by each group's $t_{p}$. On the other hand, the accumulated communication latency ($t_{m}^{\text{acc}}$) is accumulated by the maximum of $t_{p}^{\text{acc}}$ and $t_{m}^{\text{acc}}$, ensuring that the computation of the last group is finished. To address the imbalance in GEMM+A2A within MoE models, the prediction algorithm is extended by taking the maximum across all GPUs for the accumulated latencies ($t_{p}^{\text{acc}}$ and $t_{m}^{\text{acc}}$). (3) Select the optimal solution (lines 24-26). The candidate with the minimum latency is returned as the optimal solution. 

\hk{Notably, the online stage can be executed before runtime. For tasks with limited GEMM size variations (\textit{e.g.}, LLM training and text-to-video generation), the tuning is done before runtime. For dynamic tasks (\textit{e.g.}, LLM inference), we can pre-search for representative GEMM sizes, and apply nearest-neighbor matching for unseen cases during execution, eliminating the searching latency from end-to-end performance.}



\subsubsection{Tuning Stability}
\hk{We discuss the stability of the proposed tuning algorithm against thermal throttling and resource contention. The thermal throttling slows down GEMM computation but preserves the wave execution pattern. Therefore, such an impact can be handled by updating the profiled GEMM latency for predictive searching. Regarding the scenarios with predetermined resource allocation (\textit{e.g.}, preset SM ratios per process),  the predictive search can be modified to maintain accuracy from the following aspects. (1) The wave size should be determined by resource allocation. (2) GEMM and communication latencies under partial-SM configurations need updated profiling. Otherwise, if the contention pattern changes too fast or is uncontrollable, the predictive search accuracy will degrade, but the reordering technique preserves the correctness.}

\begin{table}
  \vspace{.0em}
  \caption{GEMM sizes in operator evaluation (on each GPU).}\label{tab:operator}
  \resizebox{0.48\textwidth}{!}{
  \begin{tabular}{lcccccc}
    \toprule
    \multirow{2}[0]{*}{Primitive} & \multicolumn{2}{c}{AllReduce} & \multicolumn{2}{c}{ReduceScatter} & \multicolumn{2}{c}{All-to-All} \\
     & A800 & 4090 & A800 & 4090 & A800 & 4090\\
    \midrule
    M$\times$N ($\times1024^2$) & 64$\sim$256 & 16$\sim$64 & 64$\sim$256 & 16$\sim$64 & 16$\sim$400 & 4$\sim$68\\
    K ($\times1024$) & 2$\sim$8 & 8$\sim$16 & 2$\sim$8 & 8$\sim$16 & 4$\sim$8 & 8$\sim$16\\
    \bottomrule
  \end{tabular}
  }
  \vspace{0.em}
\end{table}

\section{Implementation}
We implement \nickname based on the templated GEMMs from CUTLASS~\cite{CUTLASS}. The main loop of GEMM is well preserved, and we apply the optimal GEMM configuration tuned by the CUTLASS profiler for implementation. Following EVT~\cite{evtchen2024}, the pre-communication reordering is integrated into the epilogue of the GEMM, thereby avoiding the performance degradation of the main loop. The signaling mechanism is implemented as a GPU kernel, periodically querying the counting table to check the timing for communication. The communication is simply implemented by calling NCCL~\cite{NCCL} APIs. On top of the computation and the communication, we utilize the CUDA stream~\cite{cudastream} API to manage the concurrent execution. Specifically, the GEMM kernel is executed in a stream, while the signaling kernels and the communication kernels are executed in another stream. For each group, the signaling kernel stalls the communication until the number in the counting table meets the corresponding group size.
\section{Evaluation}
We evaluate the operator-level and end-to-end performance of \nickname with various communication primitives on different types of GPUs. The proposed design delivers 69\%-98\% of the theoretical performance and achieves up to 1.65$\times$ speedup through overlap. Furthermore, we conduct experiments to demonstrate the effectiveness of the proposed searching method and quantify the overhead of reordering. 

\begin{table}
  \vspace{.0em}
  \caption{Settings in end-to-end evaluation.}\label{tab:end2end}
  \resizebox{0.48\textwidth}{!}{
  \begin{tabular}{lccc}
    \toprule
    Application & Model & Parallelism & Setting \\
    \midrule
    LLM inference & Llama3-70B & TP=8 & \textit{chunk\_size}=16384\\
    \multirow{2}[0]{*}{LLM training} & Mixtral-8x7B & EP=4, TP=2 & \textit{input\_token}=32768\\
     & Llama3-70B & TP=8 & \textit{input\_token}=16384\\
    T2V generation & Step-Video-T2V & TP=4 & \textit{input\_token}=33792\\
    \bottomrule
  \end{tabular}
  }
  \vspace{-1.0em}
\end{table}

\subsection{Setup}
\subsubsection{Testbed}
We conduct experiments on both NVIDIA A800 GPUs and RTX 4090 GPUs. The server with A800 GPUs equips pairwise NVLink~\cite{NVLink} connecting each of the two GPUs, while the server with RTX 4090 GPUs builds inter-GPU connection traversing PCIe~\cite{PCIe} across NUMA~\cite{NUMA} nodes. The corresponding inter-GPU bandwidths are shown in Fig.~\ref{fig:curve}. The software environment includes CUDA 12.1~\cite{nvidia2025cuda}, NCCL 2.19.3~\cite{NCCL}, PyTorch 2.5.1~\cite{paszke2019pytorch}, and CUTLASS 3.6.0~\cite{CUTLASS}.

\subsubsection{Benchmark}
We benchmark GEMM+AR, GEMM+RS, and GEMM+A2A for the operator-level evaluation, with each tested under different parallelism settings and over 50 GEMM sizes from real-world workloads. The detailed sizes are presented in Tab.~\ref{tab:operator}. The end-to-end evaluation includes the scenarios of LLM inference (Llama3-70B~\cite{grattafiori2024llama3herdmodels}), LLM training (Llama3-70B~\cite{grattafiori2024llama3herdmodels} and Mixtral-8x7B~\cite{mixtral8x7b}), and text-to-video generation (Step-Video-T2V~\cite{ma2025stepvideot2vtechnicalreportpractice}). Note that the layer numbers of Llama3-70B and Mixtral-8x7B are set to 8 and 4 to fit in a single node for LLM training, respectively. The settings involved in the end-to-end evaluation are listed in Tab.~\ref{tab:end2end}. The end-to-end evaluation is conducted on the A800 GPUs.

\begin{figure}[t]
    \centering
    \includegraphics[width=0.98\linewidth]{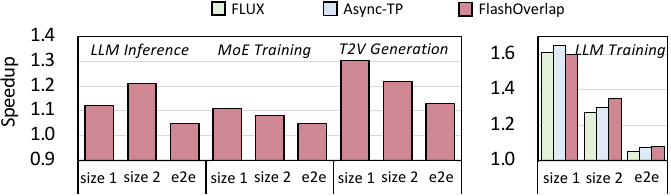}
    \vspace{-0.em}
    \caption{End-to-end ("e2e") speedup and the applied operator ("size 1" and "size 2") speedup compared to the baselines. \hk{The operator refers to the optimized "GEMM + X" part.}} 
    \vspace{-1.0em}
    \label{fig:e2e}
\end{figure}

\subsubsection{Baseline}
For the operator-level evaluation, we use both decomposition-based and fusion-based methods as baselines. One of the decomposition-based baselines is Async-TP~\cite{asynctp} by PyTorch~\cite{paszke2019pytorch}. Since Async-TP requires an NVLink connection between all GPU pairs, we further develop another decomposition-based baseline utilizing cuBLAS~\cite{cuBLAS} and NCCL~\cite{NCCL} APIs (denoted as VanillaDecomposition). For the fusion-based method, we use FLUX~\cite{fluxchang2024} as the baseline. FLUX requires peer-to-peer access, which is not supported on the tested RTX 4090 server. The non-overlap baseline is the sequential execution of GEMM and communication by calling cuBLAS and NCCL APIs, respectively. 

For the end-to-end evaluation, we use mainstream frameworks vLLM~\cite{kwon2023vllm}, Megatron-LM~\cite{megatron}, and xDiT~\cite{fang2024xdit} for LLM inference, LLM training, and text-to-video generation, respectively. We replace the original linear layer and the subsequent communication primitive with our overlap design, and use throughput as the metric for performance comparison. 

\subsection{Operator-Level Performance}
The operator-level evaluation compares the total latency of the computation and the communication. We collect the latency and calculate the speedup of each implementation normalized to the non-overlap baseline. As shown in Fig.~\ref{fig:opertor}, the speedup is the average number across multiple GEMM sizes within the corresponding range in Table~\ref{tab:operator}. Most baselines support GEMM+RS on GPUs with peer-to-peer access, as shown in Fig.~\ref{fig:opertor}(b). Compared to baselines, \nickname achieves higher average speedup, and effectively avoids performance deterioration, benefiting from interference-free computation and highly predictable overlap performance. The detailed performance comparison on typical shapes is depicted in Fig.~\ref{fig:op-a800}. Except for some cases when $K$=2048, our design consistently outperforms baselines. The exception arises from the memory access reduction of the fusion-based method such as FLUX. The memory access comprises a higher latency portion with a smaller $K$. 

\begin{figure}[t]
    \centering
    \includegraphics[width=0.98\linewidth]{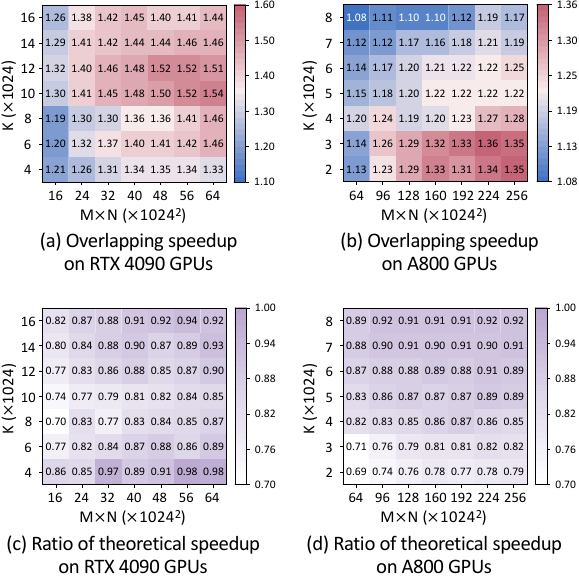}
    \vspace{-0.em}
    \caption{Performance heatmap on varying GEMM sizes. } 
    \vspace{-0.em}
    \label{fig:heatmap}
\end{figure}

Given that NVLink decreases the communication time proportion, our design yields less speedup on A800 GPUs than RTX 4090 GPUs. However, the achieved speedup ratio on the A800 GPUs relative to its theoretical speedup demonstrates a competitive result, as evidenced in Fig.~\ref{fig:heatmap} (d). On RTX 4090 GPUs, \nickname achieves 1.02-1.65$\times$ speedup over the non-overlap baseline, and 0.98-1.46$\times$ speedup over the decomposition-based method. 


\subsection{End-to-End Performance}
The end-to-end evaluation results are depicted in Fig.~\ref{fig:e2e}, and our design brings 1.05-1.13$\times$ speedup for different tasks. Since all tasks involve significant latency from GEMM and subsequent communication, the achieved speedup mainly depends on the GEMM sizes in the task. On A800 GPUs, communication overhead is relatively low. Thus, larger input sizes ($M$) and smaller intermediate sizes ($K$) yield higher speedups. Consequently, the T2V generation task benefits most from overlap due to its large input token number.  \hk{On the LLM training task, all the overlap methods demonstrate effectiveness, with \nickname gaining advantages as $K$ increases (size 2), which takes a larger latency proportion, and therefore our design achieves the highest throughput.}

\subsection{Theoretical Analysis}
The speedups across different GEMM sizes are depicted in Fig.~\ref{fig:heatmap} (a) and (b). The data on RTX 4090 GPUs and A800 GPUs are collected with GEMM+RS (TP=2) and GEMM+AR (TP=4), respectively. The number along $x$-axis is the product of $M$ and $N$ ($M\times N$), which determines the total communication data size, while the number along $y$-axis is the size of $K$ that controls the ratio between computation workload and communication data size, \textit{i.e.}, the ratio of latencies. The speedup is higher within a specific $M\times N\times K$ region, where the computation and communication latencies are closer and the acceleration space is larger. Such a region is influenced by both GPU computational capability and inter-GPU bandwidth. For example, on A800 GPUs, due to high bandwidth of NVLink, the speedup is higher with a smaller $K$.

Besides, we compare the speedup to a theoretical upper bound. Assuming the overlap is perfect, the theoretical latency is calculated by summing up the original GEMM latency and the communication latency of the final wave (if GEMM takes more time), or the GEMM latency of the first wave and the original communication latency (if communication takes more time). \nickname achieves over 80\% of the theoretical speedup in most cases. The ratio falls below 1 because of (1) under-utilized bandwidth from segmented communication (2) the prolonged computation latency brought by SM contention. The ratio is suboptimal on both GPUs with a smaller $M\times N$, as a smaller data amount, especially when segmented, fails to effectively utilize the bandwidth. 


\begin{figure}[t]
    \centering
    \includegraphics[width=0.98\linewidth]{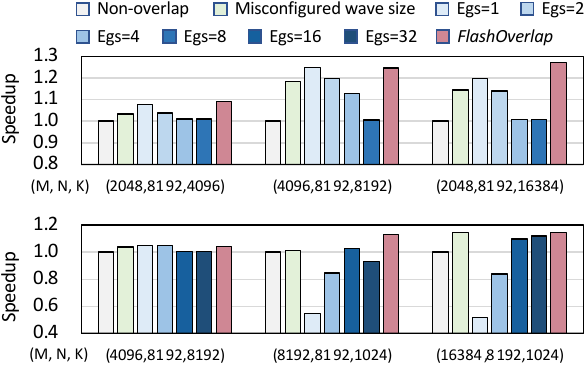}
    \vspace{-0.em}
    \caption{Performance comparison under different wave grouping strategies. Egs=$n$ denotes the equally-sized strategy with group size set to $n$ waves.} 
    \vspace{-0.0em}
    \label{fig:ablation}
\end{figure}

\subsection{Ablation Study}
\hk{We individually assess the contributions of wave grouping and the tuning algorithm. For the former, we compare the performance of \nickname against a baseline that uses a deliberately misconfigured wave size, and for the latter, we evaluate against multiple equally-sized grouping strategies. The experiments are conducted with GEMM+AR on two RTX 4090 GPUs, and with GEMM+RS on four A800 GPUs. The results are depicted in Fig.~\ref{fig:ablation} and we have the following conclusions. \textit{(1) Fixed-sized grouping fails.} For A800 GPUs, larger group sizes are preferred to mitigate bandwidth under-utilization caused by communication fragmentation. In contrast, RTX 4090 GPUs achieve better performance with a group size of 1. \textit{(2) Equally-sized grouping fails.} When communication latency dominates, larger group sizes should be adopted toward the timeline's end, as the computation finishes and no overlap space exists. \nickname outperforms all the equally-sized grouping strategies. Besides, a misconfigured wave size (+20 in the experiment) also leads to performance degradation as it introduces unavoidable communication delays of the finished tiles.} 

\begin{figure}[t]
    \centering
    \includegraphics[width=0.98\linewidth]{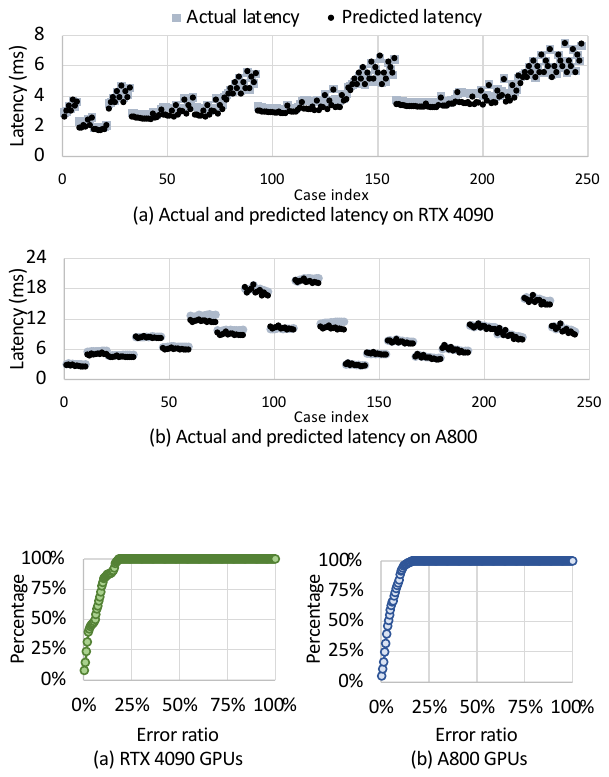}
    \vspace{-0.em}
    \caption{CDF of prediction error ratio. } 
    \vspace{-0.em}
    \label{fig:prediction}
\end{figure}

The proposed fast predictive search depends on a predictor to remove the online profiling overhead, and we measure the prediction error ratios under more than 250 combinations of different sizes, grouping partitions, and parallelism settings on each type of GPU. The cumulative distribution function (CDF) of the prediction error ratio is depicted in Fig.~\ref{fig:prediction}, and the average error ratios are 3.41\% and 3.44\% on RTX 4090 GPUs and A800 GPUs, respectively. Due to non-ideal implementation, the actual latency is always slightly higher than the predicted latency, but both demonstrate similar trends across different partitions, enabling the searched partition to be nearly optimal. To prove that, we compare the performance between the searched partition and the optimal partition. Based on such a high-accuracy predictor, the searched partitions achieve $>99\%$ performance of the optimal ones. Therefore, we directly apply the prediction-based searching in all the evaluation experiments. 

\begin{table}
  \vspace{.0em}
  \caption{\hk{Average overhead in RMSNorm and GEMM kernels. The GEMM overhead is considered to be the same for subtile-level and subtoken-level reorderings, as both are implemented as scattering operations in the epilogue.}}\label{tab:overhead}
  \resizebox{0.42\textwidth}{!}{
  \begin{tabular}{lcccc}
    \toprule
    Kernel & GPU & Tile & Subtile & Subtoken \\
    \midrule
    \multirow{2}[0]{*}{RMSNorm} & A800 & 7.46\% & 7.93\% & 8.45\% \\
    & RTX 4090 & 8.80\% & 8.78\% & 9.63\%\\
    \midrule
    \multirow{2}[0]{*}{GEMM} & A800 & 0.07\% & \multicolumn{2}{c}{0.67\%}\\
     & RTX 4090 & 0.35\% & \multicolumn{2}{c}{0.68\%}\\
    \bottomrule
  \end{tabular}
  }
  \vspace{-1.0em}
\end{table}

\subsection{Overhead Analysis} \label{sec:eval_overhead}
To quantify the overhead of reorderings, we implement an RMSNorm GPU kernel fused with the post-communication reordering operation, \hk{and measure the overhead in the epilogue of the GEMM kernel.} As mentioned in Sec.~\ref{sec:map}, the reordering granularity varies with the primitive, and the implemented RMSNorm kernel supports the reordering of tiles, subtiles, and subtokens. 
\hk{Since the reordering preserves local contiguity, its primary overhead comes from additional memory access volume of the reordering mapping table, which accounts for approximately 1.6-12.5\% of the output matrix size ($M\times N$), depending on tile size. Compared to element-wise RMSNorm, the ratio diminishes in GEMM, considering the overhead-agnostic dimension ($K$).}

\hk{The overhead evaluation covers sizes from $M$=128, $N$=1024, $K$=1024 to $M$=32768, $N$=8192, $K$=32768. 
The average overhead numbers are shown in Tab.~\ref{tab:overhead}. The post-communication reorderings of tiles, subtiles, and subtokens bring 7.46\%, 7.93\%, and 8.45\% extra latency to the RMSNorm kernel on A800 GPUs, respectively. On RTX 4090 GPUs, the numbers are 8.80\%, 8.78\%, 9.63\%, respectively. The overhead grows with smaller sizes, but the overhead remains around 10\% even for the minimal size.} 
\hk{The pre-communication reorderings are fused into the GEMM epilogue, bringing an average overhead of 0.07\%/0.35\% with tiles, and 0.67\% /0.68\% with subtiles and subtokens (both are implemented as a scattering operation) on A800/RTX 4090 GPUs.} 

\hk{\textit{Reordering pattern impact.} The subtoken-level remapping brings a relatively higher increase due to more irregular memory access, which is still inherently marginal considering the negligible latency of the element-wise operator. \textit{Matrix size impact.} Since the irregular memory access exacerbates cache line under-utilization with a small-sized matrix, the overhead grows with smaller sizes. Despite that, the overhead in RMSNorm remains around 10\% (9.2\% on A800 GPUs and 10.4\% on RTX 4090 GPUs with subtokens) even for the minimal size. For GEMM, the overhead is about 5\% (3.1\% on A800 GPUs and 5.3\% on RTX 4090 GPUs with subtokens) with extremely memory-intensive sizes, and becomes negligible as $K$ grows. \textit{Hardware impact.} GPUs with high HBM bandwidth mitigate the overhead brought by memory access. Compared to RTX 4090 GPUs, A800 GPUs have comparable Tensor Core TFLOPS (312 TFLOPS \textit{v.s.} 330 TFLOPS in FP16) but much higher HBM bandwidth (1935GB/s \textit{v.s.} 1008GB/s), yielding measurably lower overhead.}

\begin{figure}[t]
    \centering
    \includegraphics[width=0.98\linewidth]{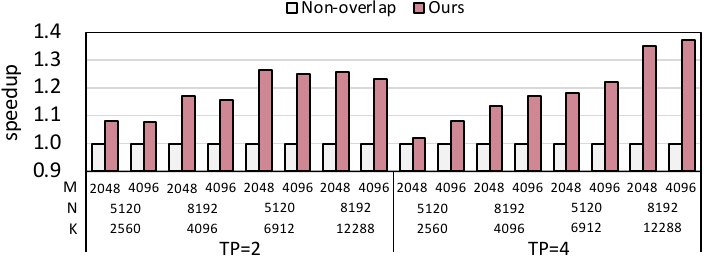}
    \vspace{-0.em}
    \caption{GEMM+AR speedup on HUAWEI Ascend NPUs. } 
    \vspace{-0.0em}
    \label{fig:huawei}
\end{figure}

\subsection{Other Platform} \label{sec:platform}
To further demonstrate the adaptability of \nickname, we implement it on HUAWEI Ascend 910B NPUs for evaluation. As for the baseline, we adopt the GEMMs from the TBE~\cite{tbe} library, which contains the operators with the state-of-the-art computational performance, and we use HCCL~\cite{hccl} for communication. The HCCL library on Ascend NPUs is analogous to the NCCL library on NVIDIA GPUs. In our design, we replace the GEMMs from TBE with templated implementations integrated with the proposed signaling mechanism. For evaluation, we choose typical GEMM shapes from LLMs, and test under TP=2 and TP=4. As depicted in Fig.~\ref{fig:huawei}, the proposed design consistently brings acceleration on all tested cases, and achieves up to 1.37$\times$ speedup for the total latency of the GEMM+AR operation. 

\section{Conclusion}
In this paper, we propose an efficient and adaptable design for data-dependent computation-communication overlap. The core idea is to utilize signals to identify the data dependency and incorporate reordering operations to enable a contiguous address for communication. We design the signaling timing to enhance efficiency and extend it to be tunable. For real-time searching, we introduce a predictive search method. The experiments demonstrate the efficacy of the proposed design, showing a speedup of up to 1.65$\times$ through overlap.

\begin{acks}
\hk{We sincerely thank our shepherd, Yida Wang, and anonymous reviewers for their feedback and insightful suggestions. This work is supported by the National Natural
Science Foundation of China (No. 62325405, U24B6015), Beijing Natural Science Foundation (No. L242018, L257010), Beijing Municipal Science and Technology Project (No. Z241100004224013), Beijing National Research Center for Information Science, Technology (BNRist), Beijing Innovation Center for Future Chips, and State Key Laboratory of Space Network and Communications.} 
\end{acks}

\bibliographystyle{ACM-Reference-Format}
\bibliography{my}

%
\newpage
\appendix
\section{Artifact Appendix} 
This appendix describes the artifact evaluation of the paper \textit{Efficient and Adaptable Overlapping for Computation and Communication via Signaling and Reordering}.

\subsection{Abstract}
The paper \textit{Efficient and Adaptable Overlapping for Computation and Communication via Signaling and Reordering} proposes a novel computation-communication overlap method, named \textit{FlashOverlap}. The following artifacts demonstrates the functionality and efficiency of \textit{FlashOverlap}, including three experiments corresponding to the three major claims in the paper: (1) The correctness and speedup of \textit{FlashOverlap}. (2) The accuracy of the predictive search method in determining the overlap configuration. (3) The negligible overhead brought by the design via kernel fusion. 

\subsection{Description \& Requirements}

\subsubsection{How to access}
The source code is available at the \textit{ae} branch in the public GitHub repository:

\url{https://github.com/infinigence/FlashOverlap/tree/ae},
and published on Zenodo:

\url{https://doi.org/10.5281/zenodo.17201530}.

\subsubsection{Hardware dependencies}
The artifacts require \textit{sm80}, \textit{sm86}, \textit{sm89} NVIDIA GPUs to run, and the experiments described in the paper are conducted on A800 GPUs and RTX 4090 GPUs. The codes can also be used on RTX 3090 and A100 GPUs, but the results are not reported. 

\subsubsection{Software dependencies}
\begin{itemize}
    \item CUDA 12.1, 12.2 (version not mandated)
    \item PyTorch 2.7.0 (version not mandated)
    \item CUTLASS $\geq$ 3.6.0, $\leq$ 3.9.0
    \item NCCL 2.18.3, 2.19.3 (version not mandated)
    \item cmake $\geq$ 3.18
\end{itemize}

\subsubsection{Benchmarks} 
None. We use randomly generated inputs for both correctness and performance evaluation. 

\subsection{Setup}\label{sec:ap_setup}
Please follow the \textbf{README.md} in the repository for installation. Before evaluation, we need to first profile the General Matrix Multiplication (GEMM), and then tune the overlap configurations. For convenience, we provide a unified script to execute profiling and tuning across multiple cases, simply by running \textit{evaluation/preparation.py}.

\subsection{Major Claims}
\subsubsection{Claim (C1)}: Our design maintains the mathematical equivalence with the non-overlap implementation, while delivering up to 1.65$\times$ speedups through overlap. The supported communication primitives include AllReduce, ReduceScatter, and All-to-All, and we denote the corresponding overlap patterns as GEMM+AR, GEMM+Reduce, and GEMM+A2A, respectively. This is proven by the experiment (E1) described in Section~6.2 whose results are reported in Figure~10.

\subsubsection{Claim (C2)}: The proposed predictive search method achieves > 99\% performance of the optimal ones searched by the exhaustive method, with the average error ratio of the predictor remaining below 5\%. This is proven by the experiment (E2) described in Section~6.5 whose results are reported in Figure~14.

\subsubsection{Claim (C3)}: Our design brings negligible overhead to both the GEMM kernel and the subsequent element-wise kernel (\textit{e.g.}, RMSNorm), with measured overheads kept within 1\% and 10\%, respectively. This is proven by the experiment (E3) and results described in Section~6.6.

\subsection{Experiments}

\subsubsection{Experiment (E1): Overlap performance.} \textit{[1 human-hour + 3h compute-hour]}. The experiment evaluates the correctness of our method against the non-overlap design, and shows the overlap speedup of GEMM+AR, GEMM+RS, and GEMM+A2A. Besides, on GPUs with peer-to-peer access (\textit{e.g.}, A800 GPUs), the experiment also compares the GEMM+RS performance of our design and SOTA implementations.

\textit{[Preparation].}
Follow the steps in \S\ref{sec:ap_setup} for setup, and make sure the configurations are generated. 

\textit{[Execution].}
(1) Open the \textbf{evaluation} directory, and run \textit{e1\_correctness.py}. (2) Run \textit{e1\_speedup.py}. (3) On GPUs with P2P access, run \textit{e1\_compare.py}. 

\textit{[Results].}
(1) The terminal outputs \textit{all close} for 10 randomly selected cases. 
(2) The terminal outputs a table listing the speedups on 2,4,8 GPUs and for all the primitives (up to 1.30$\times$ on A800 GPUs and 1.65$\times$ on RTX 4090 GPUs). (3) The terminal outputs the speedups against SOTA implementations for GEMM+RS, across different GPU numbers (\textit{FlashOverlap} is slightly better).


\subsubsection{Experiment (E2): Search accuracy.} \textit{[1 human-hour + 5 compute-hour]}. The experiment quantifies the latency prediction error and compares the predictive search with an exhaustive search baseline, therefore being time-consuming due to the exhaustive search.

\textit{[Preparation].}
Follow the steps in \S\ref{sec:ap_setup} for setup, and make sure the configurations are generated. 

\textit{[Execution].}
Open the \textbf{evaluation} directory, and run the script \textit{e2\_predictive\_search.py}.

\textit{[Results].}
The terminal outputs (1) the cumulative distribution curve of the average prediction error and (2) the performance comparison between the predictive searched solution and the exhaustive searched solution (>99\%).

\subsubsection{Experiment (E3): Overhead.} [10 human-minutes + 30 compute-minutes]: The experiment quantifies the overhead fused into the GEMM kernel and the RMSNorm kernel.

\textit{[Preparation].}
Follow the steps in \S\ref{sec:ap_setup} for setup, and make sure the configurations are generated.

\textit{[Execution].}
(1) Open the \textbf{evaluation} directory, and run \textit{e3\_rmsnorm\_overhead.py}. (2) Run \textit{e3\_gemm\_overhead.py}. 

\textit{[Results].}
(1) The terminal outputs a table containing the average overhead ratio under three different patterns (within 10\%). (2) The terminal outputs a table containing the average overhead ratio under two different patterns (within 1\%). 

\subsection{Notes on Reusability}
\subsubsection{More platforms} 
When adapting to more hardware platforms, the dominant efforts lie in the utilization of the templated GEMM implementation (\textit{e.g.}, CUTLASS~\cite{CUTLASS}). On Hopper GPUs, new GEMM templated configurations, such as thread block cluster shape, are introduced. Therefore, the GEMM configuration profiling scripts necessitate modification. On the other hand, since the proposed design implements communication via API calls, the communication component requires minimal modifications when adapting to other platforms. 
\subsubsection{Inter-node Communication}
For the current implementation, we use multi-processing for communication, which is limited to intra-node parallelism. Therefore, when adapting to the inter-node communication, we ought to switch to the distributed communication package provided by PyTorch, whereas the backends for the GEMM component and the communication remain the same.

\end{document}